\documentstyle[12pt]{article} 
\begin{document}                                                              
\begin{titlepage}
\begin{flushright}
BNL-GGP-2\\
November, 1993\\
\end{flushright}
\vfill
\begin{center} 
{\Large\bf QCD and the chiral critical point}\\
\vfill
{\large \bf Sean Gavin, Andreas Gocksch, \& Robert D. Pisarski}\\
{\large Department of Physics\\
Brookhaven National Laboratory\\
P.O. Box 5000\\
Upton, New York  11973-5000} \\ 
\vfill
{\bf Abstract}
\end{center}
\begin{quotation}

As an extension of $QCD$, consider a theory with ``$2+1$'' flavors,
where the current quark masses are held in a
fixed ratio as the overall scale of the quark masses is varied.
At nonzero temperature and baryon density it is expected that in the
chiral limit the chiral phase transition is of first order. 
Increasing the quark mass from zero,
the chiral transition becomes more weakly first order, and can end in a chiral
critical point.  We show that the only massless field at the chiral
critical point is a sigma meson, with the
universality class that of the Ising model.  Present day lattice simulations
indicate that $QCD$ is (relatively) near to the chiral critical point.

\end{quotation}
\vfill
\end{titlepage}

Understanding the collisions of heavy ions at ultrarelativistic energies
requires a detailed knowledge of the equilibrium
phase diagram for $QCD$ 
at nonzero temperature and baryon density.
We generalize $QCD$ to a nonabelian gauge theory 
with three colors and ``$2+1$'' flavors
by holding the current quark masses 
in a fixed ratio as the overall mass scale is varied:
$m \equiv m_{up} = m_{down} = r \, m_{strange}$, 
with $r$ a constant of order $\sim 1/20$.  (For our purposes
the difference between the up and down quark masses is inconsequential.)
Currently, numerical simulations of lattice gauge theory [\ref{rev}]
find that while there are lines
of first order transitions coming up from $m=0$ and down from $m=\infty$,
these lines do {\it not} meet --- there is a gap, with $QCD$ somewhere
in between.  This is illustrated in fig. (1), following a similar diagram
from the results of the Columbia group [\ref{col}].

Lines of first order transitions typically end in critical points, so it
is natural to ask about the two critical points, labeled
``${\cal C}$'' and ``${\cal D}$'' in fig. (1).
As $m$ decreases from $m=\infty$, 
the line of deconfining first order phase transitions [\ref{ys}]
can end in a deconfining critical point, ``${\cal D}$'' in
fig. (1).
Correlation functions between Polyakov lines are infinite ranged at
the deconfining critical point; by an analysis similar to that in
(II) below, one can show that ${\cal D}$ lies
in the universality class of the Ising model, or a $Z(2)$
spin system, in three dimensions.

The opposite limit is to work up from zero quark mass.  For three flavors
the chiral phase transition is expected to be
of first order at $m=0$ [\ref{pw},\ref{gold}], so as $m$ increases,
the line of first order transitions can end in a chiral critical point,
``${\cal C}$'' in fig. (1).
In this Rapid Communication we show that for $2+1$ flavors
there is only one massless field at the chiral critical point,
a sigma meson ($J^P = 0^+$, predominantly isosinglet);
the universality class is again that of the Ising model.
Notice, however, that very different fields go critical at the two
critical points ${\cal C}$ and ${\cal D}$.

{\bf I.}  We start at zero temperature by
fitting the scalar and pseudoscalar mass spectrum in $QCD$
to that found in a linear sigma model [\ref{pw}-\ref{chan}].
For three quark flavors we introduce the field $\Phi$ 
($\sim \overline{q}_{left} \, q_{right}$), as a complex valued,
three by three matrix, 
$ \Phi = \sum^{8}_{a=0} ( \sigma_a + i \, \pi_a ) t^a$; 
$\, t_{1 \ldots 8}$ are the generators of the $SU(3)$ algebra in
the fundamental representation, and $t_0$ is proportional to the
unit matrix.  Normalizing the generators as $tr(t_a t_b) = \delta^{a b}/2$,
$t_0 = {\bf 1}/\sqrt{6}$.

The fields $\sigma_a$ are components of a
scalar ($J^P = 0^+$) nonet, those of $\pi_a$ a pseudoscalar 
($J^P = 0^-$) nonet.  The latter are familiar, as
$\pi_{1,2,3}$ are the three pions, denoted as $\pi$ without
subscript, and the
$\pi_{4,5,6,7}$ are the four kaons, the $K$'s.
The $\pi_8$ and $\pi_0$ mix to form the 
mass eigenstates of the $\eta$ and $\eta'$ mesons, with mixing
angle $\theta_{\eta \eta'}$ [\ref{pdt}].
For notational ease we define the components of the scalar
nonet analogously: we refer to $\sigma_{1,2,3}$ as the $\sigma_\pi$'s, 
to $\sigma_{4,5,6,7}$ as the $\sigma_K$'s, while $\sigma_8$ and $\sigma_0$
mix to form the $\sigma_\eta$ and $\sigma_{\eta'}$.
This multiplicity of eighteen fields is to be contrasted with the
usual sigma model with two flavors, which only has three $\pi$'s and
one $\sigma$ meson.  

The effective lagrangian for the $\Phi$ field is taken to be
[\ref{pw}-\ref{chan}]
\begin{equation}
{\cal L} \; = \; tr \left| \partial_\mu \Phi \right|^2
\; - \; tr \left( H (\Phi + \Phi^\dagger) \right)
\; + \;  \mu^2 \; tr\left(\Phi^\dagger \Phi\right) 
\label{e2}
\end{equation}
$$
\; - \; \sqrt{6} \; c \left( det(\Phi) \, + \, det(\Phi^\dagger) \right)
\; + \; (g_1 - g_2) \left( tr \Phi^\dagger \Phi \right)^2
\; + \; 3 \, g_2 \; tr\left(\Phi^\dagger \Phi\right)^2 \;
$$
The parameters of the linear sigma model are
the background field $H$, a mass parameter $\mu^2$, an ``instanton''
coupling constant $c$, and two quartic couplings, $g_1$ and $g_2$.
Neglecting effects from $m_{up}\neq m_{down}$, 
for the background field $H$ we take 
$H = h_0 \, t_0 - \sqrt{2} \, h_8 \, t_8$.  The current
quark masses are then related to the background field as
$m_{up} = m_{down} \sim h_0 - \, h_8$ and 
$m_{strange} \sim h_0 + 2 \, h_8$.

Throughout this paper we work exclusively at the simplest level
of mean field theory.  Up to differences in normalization
our intermediate results agree with those of Chan and Haymaker [\ref{chan}].
Because we attempt to fit to current experimental results [\ref{pdt}],
our fit in (\ref{e11}) differs from theirs;
details will be presented elsewhere [\ref{ggp}].

We assume that 
there are nonzero vacuum expectation values for $\sigma_0$
and $\sigma_8$,
$ \sigma_0  \rightarrow  \Sigma_0 + \sigma_0 $,
$ \sigma_8 \rightarrow - \, \sqrt{2} \, \Sigma_8 + \sigma_8 $.
Expanding the lagrangian in powers of $\sigma_a$ and $\pi_a$,
expansion to linear order fixes the values of 
$\Sigma_0$ and $\Sigma_8$, while expansion to quadratic order gives 
the masses of all the fields: the mass of the pion, $m_\pi$, {\it etc.}

There is one unexpected feature of the results.
For the two equations of motion,
the masses of the entire pseudoscalar nonet
(for $m_\pi$, $m_K$, $m_\eta$, and $m_{\eta'}$),
and the masses of 
half the scalar nonet (for $m_{\sigma_{\pi}}$ and $m_{\sigma_{K}}$),
the two quantities $\mu^2$ and $g_1$ 
only enter in tandem, through the new parameter
$ M^2 = \mu^2 + g_1 \, (\Sigma^2_0 + \, 2 \, \Sigma^2_8)$.
This means that we can fit to the pseudoscalar spectrum, and so fix $M^2$,
and yet still be free to vary $g_1$: the {\it only}
change is to alter the masses of the $\sigma_{\eta}$ and the
$\sigma_{\eta'}$.  This technical detail plays an important role in
what follows; although
there must be some simple group theoretic reason for it,
as of yet we do not know what it is.

We determine the parameters of the linear sigma model by fitting to 
$m_\pi = 137 \, MeV$,
$m_\eta = 547 \, MeV$, and
$m_{\eta'} = 958 \, MeV$, and to the pion decay constant,
$f_{\pi} = 93 \, MeV$.
A solution is 
$$
\Sigma_0 \; = \; 127 \, MeV \;\; , \;\;
\Sigma_8 \; = \; 13 \, MeV \;\; , \;\;
h_0 \; = \; (290 \, MeV)^3 \;\; , \;\;
h_8 \; = \; (281 \, MeV)^3 \;\; , \;\;
$$
\begin{equation}
M^2 \; = \; + \, (642 \, MeV)^2 \;\; , \;\;
c \; = \; 1920 \, MeV \;\; , \;\;
g_2 \; = \; 30 \; .
\label{e11}
\end{equation}
Following [\ref{chan}], these values were obtained by
varying $\Sigma_8/\Sigma_0$ as a free parameter.
This is possible because 
the kaon mass is rather insensitive to $\Sigma_8/\Sigma_0$:
for (\ref{e11}) the kaon mass is
a bit high, $m_K^{fit} = 516 \, MeV$ instead of the (average) 
experimental value of $497 \, MeV$.  Other quantities, however, 
are most sensitive to $\Sigma_8/\Sigma_0$; in
particular, the kaon decay constant favors a small ratio,
as $f_K^{fit} = 109 \, MeV$ is 
close to the experimental value of $113 \, MeV$.  Also,
the $\eta$-$\eta'$ mixing angle, $\theta^{fit}_{\eta \eta'} = - \, 10.4^{o}$,
is reasonable.  (The value obtained from radiative decays [\ref{gil}],
$\theta^{rad}_{\eta \eta'} = - \, 20^{o}$, favors even smaller values of
$\Sigma_8/\Sigma_0$.)
Further, because $\Sigma_8 \neq 0$, the 
ratio of the strange to up ($=$ down) quark masses is 
$m_{strange}/m_{up} = (h_0 + 2 h_8)/(h_0 - h_8) = 32$, and not
the often quoted value of $\sim 20$.

The fit makes unique predictions for two masses in the scalar
nonet, 
$m_{\sigma_\pi} = 1177 \, MeV$ and
$m_{\sigma_K} = 1322 \, MeV$.
There are observed states [\ref{pdt}] with these quantum numbers, the
$a_0(980)$ and the $K^*_0(1430)$, respectively; the values for the
$\sigma_\pi$ and the $\sigma_K$ aren't too far off, although the splitting
between them is too small.
We note that the 
identification of the $a_0$ with the $\sigma_\pi$ is problematic
(VII.21 of ref. [\ref{pdt}]): the $a_0(980)$ may not be the 
$\sigma_\pi$ [\ref{torn}], but a $K \overline{K}$ molecule [\ref{kkbar}].  

There is no unique prediction for two other members of the scalar
nonet, the $\sigma_{\eta}$ and the $\sigma_{\eta'}$.   As remarked,
the masses of all other fields only depends upon the 
parameter $M^2$.  In fig. (2) we illustrate 
how $m_{\sigma_{\eta}}$ and $m_{\sigma_{\eta'}}$ change as $g_1$ is
varied at fixed $M^2=+(642 \, MeV)^2$.
We identify the $\sigma_{\eta'}$ and the $\sigma_{\eta}$ with the
observed states [\ref{pdt}] with the same quantum numbers: the $f_0(975)$
and the $f_0(1400)$, respectively.  With the parameters of (\ref{e11}),
if we require that $m_{\sigma_{\eta '}} = 975 \, MeV$, fig. (2) predicts that
$m_{\sigma_{\eta}} = 1476 \, MeV$ instead of $1400 \, MeV$.  Also,
$g_1 = 40$, $\mu^2 =  - \, (492 \, MeV)^2$, and
the mixing angle between the $\sigma_{\eta}$ and the $\sigma_{\eta'}$
is $+ 28^o$.
As before the identification of the $\sigma_{\eta '}$ with the
$f_0(975)$ is open to question (VII.192 of ref. [\ref{pdt}]):
the $f_0(975)$ may be not the $\sigma_{\eta'}$ [\ref{penn}], 
but a $K\overline{K}$ molecule [\ref{kkbar}].
For the analysis of (III), all that is important is that
the $\sigma_{\eta'}$ is not light [\ref{com1}], so at zero
temperature the quartic coupling $g_1$ is large.

{\bf II.} The details of the spectrum at zero temperature are not 
needed to understand how a chiral critical point can arise.  
In mean field theory the effects of nonzero
temperature or baryon density are incorporated simply by varying the
mass parameter $\mu^2$.  This is valid in the limit of very high
temperature, but should be qualitatively correct at all temperatures.  

We begin with the $SU(3)$ symmetric case, $h_8 = 0$.  For a constant
field $\Sigma_0$ the lagrangian reduces to the potential for $\Sigma_0$, 
\begin{equation}
{\cal L} \; = \; - h_0 \, \Sigma_0 
\; + \; \frac{1}{2} \, \mu^2 \, \Sigma_0^2
\; - \; \frac{c}{3} \, \Sigma^3_0
\; + \; \frac{g_1}{4} \, \Sigma^4_0 \; .
\label{e15}
\end{equation}
This model has precisely the same phase diagram as that for the 
phase transition
between a liquid and a gas.  For zero background field, $h_0 = 0$, the
instanton interaction $det(\Phi) \sim \Sigma_0^3$ is cubic and 
and so drives the transition
first order.  As $h_0$ increases the transition becomes
more weakly first order, until at $h_0 = h_0^{crit}$
the line of first order transitions ends
in a critical point.  For $h_0 > h_0^{crit}$ there is no
true phase transition, just a smooth crossover.

The critical point occurs when
$h_0^{crit} = c^3/(27 \, g^2_1)$, 
$\Sigma_0^{crit} = c/(2 \, g_1)$, and
$\mu^2_{crit} = c^2/(3 \, g_1)$.
At this point the 
potential in $\Sigma_0 - \Sigma_0^{crit}$
is purely quartic, ${\cal L} = g_1 (\Sigma_0 - \Sigma_0^{crit})^4/4$, so
$m^2_{\sigma_{\eta'}} = 0$.
The other fields are all massive:
$m_\pi^2 = m_K = m_\eta^2 = c^2/(9 g_1)$, $m^2_{\eta '} = 10 m^2_\pi$, 
$m_{\sigma_{\pi}}^2 = m_{\sigma_{K}}^2 = m_{\sigma_{\eta}}^2 =
(7 + 18 g_2/g_1) m^2_\pi$.  
Since only the $\sigma_{\eta'}$ is massless at the chiral critical point,
the similarity to the liquid
gas phase transition extends to the universality class, which is
that of the Ising model.

This conclusion remains true away from the case of $SU(3)$ symmetry, 
$h_8 \neq 0$.  Numerical analysis [\ref{ggp}] shows that 
there is a single, massless field
at the chiral critical point, the $\sigma_{\eta '}$,
with the universality class that of the Ising model.  Of course for
$h_8 \neq 0$ the $\sigma_{\eta'}$ field does not
remain a pure $SU(3)$ singlet, but mixes to become part octet.  

The possibility of a chiral critical point can even be seen from the
calculation of the zero temperature spectrum in fig. (2).  Although we
did not remark upon it before, when the coupling 
$g_1 \sim 3.8$, $m_{\sigma_{\eta'}} = 0$.  There it appears
as mere curiosity; after all, in fig. (2) $\mu^2$ has the value appropriate
to zero temperature, while the value of $\mu^2$ 
at nonzero temperature (or baryon density) must be larger.  Even so, 
fig. (2) does illustrate how a single field, the $\sigma_{\eta '}$,
can become massless at a special point in the phase diagram.

We can further explain the nature of the entire phase diagram as a function
of $m_{up} = m_{down}$ versus $m_{strange}$, as proposed in fig. (1)
of ref. [\ref{col}].  The basic idea is along the entire line of
chiral critical points, only the $\sigma_{\eta'}$ is massless, but that
the singlet/octet ratio in the $\sigma_{\eta'}$ changes.  At the
$SU(3)$ symmetric point, 
$m_{up}=m_{down}=m_{strange}$, the $\sigma_{\eta'}$
is an $SU(3)$ singlet.  Decreasing $m_{strange}$ to the critical point where
$m_{strange} =0$ and 
$m_{up}=m_{down} \neq 0$, the $\sigma_{\eta'}$ becomes
entirely strange,
$\sigma_{\eta'} \sim \overline{s} s$.  

The opposite limit of $SU(2)$ chiral symmetry, 
$m_{up}=m_{down}=0$, is more familiar.
Assume, as in ref. [\ref{col}], that the chiral phase transition with
two massless flavors is of second order, with the universality class 
that of an $O(4)$ critical point.
There is then a special value of 
$m_{strange} = m_{strange}^{crit}$, with a
line of $O(4)$ critical points for $m_{strange} > m_{strange}^{crit}$, 
and a line of first order transitions when
$m_{strange} < m_{strange}^{crit}$.
Wilczek [\ref{wil}] observed that exactly at
$m_{strange} = m_{strange}^{crit}$, the chiral transition is in
the universality class of an $O(4)$ {\it tri}critical point.  
In our view, at the critical points along
$m_{strange} \geq m_{strange}^{crit}$,
the $\sigma_{\eta '}$ is a pure
$SU(2)$ state, with $\sigma_{\eta '} \sim  \overline{u} u + \overline{d} d$,
while the pions are massless because $m_{up}=m_{down}=0$.
Whether the universality class is $O(4)$ critical or $O(4)$ 
tricritical depends upon the relevant quartic couplings.

The analysis can be extended to the case where 
the chiral transition is of first order for two, massless flavors,
but the gap between chiral and deconfining regions remains.
Then fig. (1) of ref. [\ref{col}] would have to 
be modified, with a band of first
order transitions about the axis $m_{up} = m_{down} = 0$.  
These first order transitions would end in chiral critical points,
in the Ising universality class from the presence of 
massless $\sigma_{\eta'}$ fields.

{\bf III.} Lastly we ask: 
{\it How far is $QCD$ from the chiral critical point?}
In fig. (1) we have grossly
exaggerated the case, putting $QCD$ very
close to the chiral critical point.  But the data of ref. [\ref{col}] 
does indicate that as a function of $m$,
$QCD$ is only about a factor of two from the chiral critical point.

We first {\it assume} that the chiral phase transition is of first order
for three, massless flavors because of the presence of the instanton
coupling $\sim det(\Phi)$.  
To find the chiral critical point we vary the current
quark masses, or equivalently the background fields $h_0$ and
$h_8$.  We require that the ratio
of strange to up quark masses equals the value found from the fit
at zero temperature, $(h_0 + 2 h_8)/(h_0 - h_8)=32$.
By varying $h_0$, $h_8$, $\mu^2$, $\Sigma_0$, and
$\Sigma_8$, and otherwise using the values found in (\ref{e11}) 
we find that the critical point occurs for 
$h_0^{crit} = (62 \, MeV)^3$,
$h_8^{crit} = (60.4 \, MeV)^3$,
$\mu^2_{crit} = (183 \, MeV)^2$, 
$\Sigma_0^{crit} = 14.5 \, MeV$, and
$\Sigma_8^{crit} = 2.7 \, MeV$.  
We then compute the ratio between the current quark masses at the
chiral critical point to those in $QCD$:
\begin{equation}
\frac{m^{crit}_{up}}{m_{up}} \; = \;
\frac{h_0^{crit} - h_8^{crit}}{h_0 - h_8} \; = \;
.01 \; .
\label{e17}
\end{equation}
Now we readily confess that this ratio, computed in
mean field theory, is at best crude.  Even so, numerical 
simulations [\ref{col}] find that the ratio in (\ref{e17}) 
is not $.01$, but $\sim .5$ --- 
mean field theory is off by almost two orders of magnitude!

Thus our initial assumption must be false.
In the chiral limit for
three flavors, there are two mechanisms for generating a
first order transition.  The first is the presence of the 
instanton coupling $\sim det(\Phi)$
[\ref{pw},\ref{gold}].  From (\ref{e17}) this 
gives the wrong phase diagram.
Therefore the second mechansim must be operative, a type of
Coleman-Weinberg transition [\ref{cw}].

In mean field theory quartic couplings are fixed and do not change with
temperature; a Coleman-Weinberg transition is one in which
the quartic couplings run from a stable into an unstable regime,
and so thereby generate a finite correlation length dynamically.
This phenomenon can be demonstrated rigorously in $4$ and $4-\epsilon$
dimensions [\ref{cw}-\ref{eps}];
extrapolation to three dimensions, $\epsilon = 1$, is open to question.

If in the chiral limit the phase transition is of first order because
it is Coleman-Weinberg, it is reasonable to suggest that even for
nonzero quark mass the quartic couplings change significantly with
temperature.  Since $QCD$ appears to be in a region of smooth crossover,
presumably the quartic couplings of the sigma model
do not run all the way into the unstable regime.
This could 
explain the apparent discrepancy with mean field theory in (\ref{e17}).
At the very least
the quartic couplings do tend to run in the right direction:
in the infrared limit in less than four dimensions [\ref{shen},\ref{eps}], 
the couplings run most strongly for large $g_2$, 
approximately at constant $g_2$ from large to small values of $g_1$.
This type of running is {\it precisely} 
what is required at the chiral critical
point: as illustrated in fig. (2), even at zero temperature we can
make $\sigma_{\eta'}$ massless by going from large to small values
of $g_1$, keeping $g_2$ fixed. 

Given the limitations inherent in present lattice simulations
for $2+1$ flavors, we conclude with a
conjecture: as suggested by our illustration in fig. (2), perhaps
$QCD$ is {\it very} close to the chiral critical point.  
There is no good reason why it should be;
but if we are lucky, then even if there is no true phase transition in
$QCD$, the $\sigma_{\eta '}$ could still become very light.
As we alluded to previously [\ref{dcc}], a light $\sigma_{\eta'}$
may generate large domains of Disoriented Chiral Condenstates
in the collisions of large nuclei at ultrarelativistic energies [\ref{ggp}].  

Speculation aside, we have shown how 
a detailed understanding of the spectrum of the scalar nonet 
at zero temperature ---
especially knowing exactly 
where the $\sigma_{\eta'}$ meson lies --- has direct and
dramatic consequences for the phase diagram of $QCD$ at nonzero
temperature and baryon density.  
In this way, one area of hadronic physics unexpectedly
fertilizes another.

R.D.P. thanks Carl Dover, Yue Shen, and Edward Shuryak for discussions.  
This work was supported by a DOE grant at 
Brookhaven National Laboratory (DE-AC02-76CH00016) and by
a National Science Foundation grant at the Institute for Theoretical
Physics (PHY89-04035).

\vspace{.25in}
\noindent{ \bf References}
\newcounter{nom}
\begin{list}{[\arabic{nom}]}{\usecounter{nom}}
\item
F. Karsch and E. Laermann, Bielefeld preprint BI-TP 93-10
(March, 1993), hep-lat/9304010, and references therein.
\label{rev}
\item
F. R. Brown {\it et al.}, Phys. Rev. Lett. 65, 2491 (1990).
\label{col}
\item
B. Svetitsky and L. G. Yaffe, Nucl. Phys. B210, 423 (1982).
\label{ys}
\item
R. D. Pisarski and F. Wilczek, Phys. Rev. D29, 338 (1984).
\label{pw}
\item
H. Goldberg, Phys. Lett. B131, 133 (1983).
\label{gold}
\item
L.-H. Chan and R. W. Haymaker, Phys. Rev. D7, 402 (1973).
\label{chan}
\item
Review of Particle Properties, Particle Data Group, Phys. Rev. D45, S1 (1992).
\label{pdt}
\item
S. Gavin, A. Gocksch, and R. D. Pisarski, work in progress.
\label{ggp}
\item
F. J. Gilman and R. Kauffman, Phys. Rev. D36, 2761 (1987); (E)
D37, 3348 (1988).
\label{gil}
\item
N. A. Tornqvist, Nucl. Phys. B (Proc. Suppl.) 21, 196 (1991), 
and references therein.
\label{torn}
\item
J. Weinstein and N. Isgur, Phys. Rev. D41, 2236 (1990), 
and references therein.
\label{kkbar}
\item
M. Pennington, Nucl. Phys. B (Proc. Suppl.) 21, 37 (1991), 
and references therein.
\label{penn}
\item
We recognize that models of nucleon-nucleon
forces do postulate the existence of a 
$\sigma$ meson with a mass of $\sim 600 \, MeV$.  It is reasonable
to view this $\sigma$ meson as an approximation to two pion exchange;
there is no such resonance seen
in the phase shifts of $\pi - \pi$ scattering below $1\, GeV$,
VII.37 of ref. [\ref{pdt}].  For the purposes of discussion, we note that
if we take $m_{\sigma_{\eta'}} = 600 MeV$, the coupling $g_1$ decreases
to $g_1 = 16.2$, while $m_{\sigma_{\eta}} = 1412 MeV$ and 
$\mu^2 = + (384 MeV)^2$.  The ratio in (4) goes up, from $.01$ to
$.06$, which is still a long way from $\sim .5$, ref. [\ref{col}].
\label{com1}
\item
F. Wilczek, Intern. J. Mod. Phys. A7, 3911 (1992);
K. Rajagopal and F. Wilczek, Nucl. Phys. B399, 395 (1993).
\label{wil}
\item
D. J. Amit, ``Field Theory, the Renormalization Group, and Critical
Phenomenon,'' (World Scientific, Singapore, 1984), sec. II.4, and
references therein.
\label{cw}
\item
R. S. Chivukula, M. Golden, and E. H. Simmons, 
Phys. Rev. Lett. 70, 1587 (1993).
\label{cgs}
\item
Y. Shen, Phys. Lett. B315, 146 (1993); 
Boston University preprint BUHEP-93-24, hep-lat/9311018.
\label{shen}
\item
A. J. Paterson, Nucl. Phys. B190, 188 (1981);
R. D. Pisarski and D. L. Stein, Phys. Rev. B23, 3549 (1981);
J. Phys. A14, 3341 (1981). 
\label{eps}
\item
S. Gavin, A. Gocksch, and R. D. Pisarski, 
Brookhaven preprint BNL-GGP-1 (Oct., 1993), hep-ph/9310228.
\label{dcc}
\end{list}

\vspace{.25in}

\noindent { \bf Figure Captions} 
\vspace{.1in}

\noindent {\it Fig. 1:} Proposed phase diagram for $2+1$ flavors, 
following ref. [2]: ${\cal C}$ is the chiral critical point, 
${\cal D}$ the deconfining critical point.\\

\noindent {\it Fig. 2:} Plot of the masses of the 
$\sigma_{\eta}$ and the $\sigma_{\eta'}$
versus the coupling $g_1$ for the fit of (2).
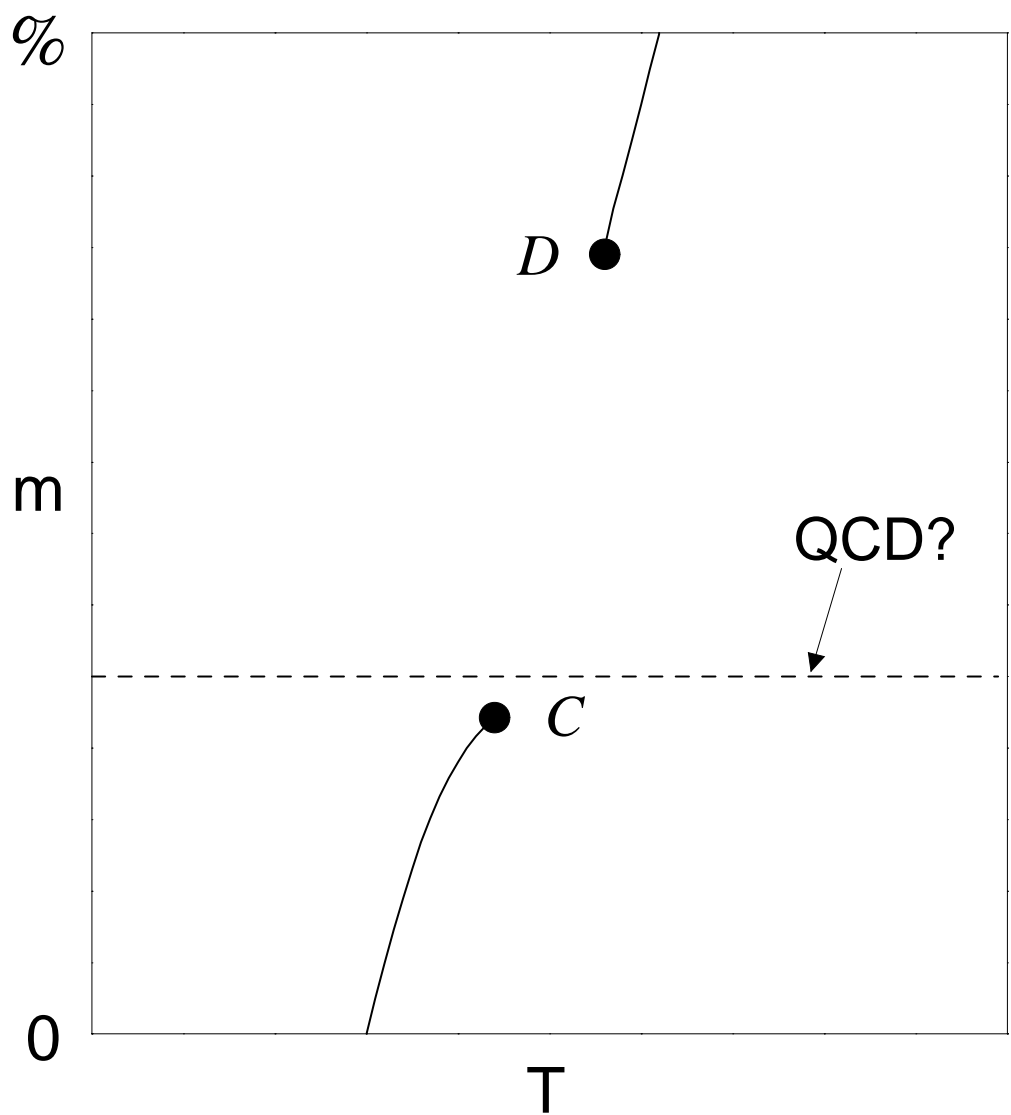
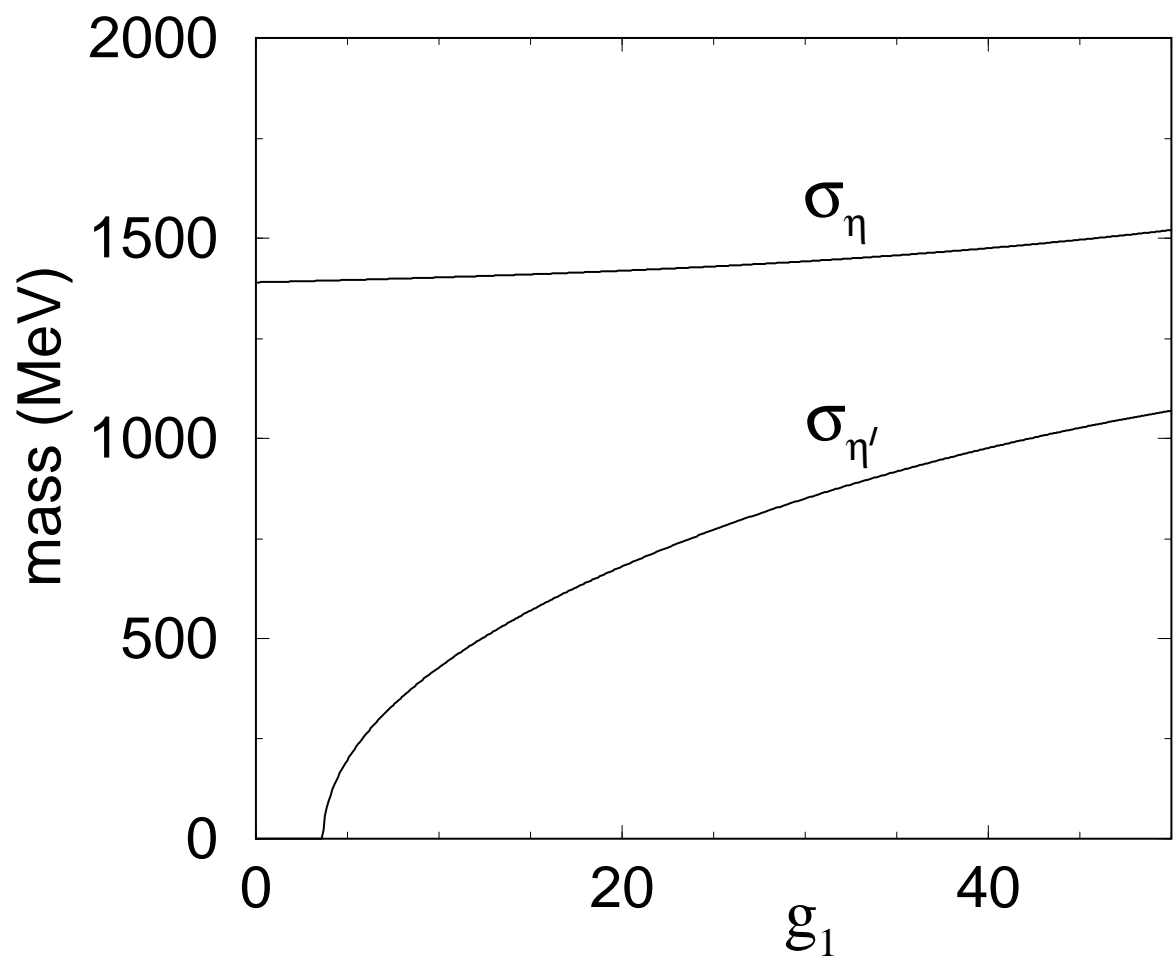
\end{document}